\documentclass[12pt,article,onecolumn]{IEEEtran}
\usepackage{amsmath}
\usepackage{amssymb}
\usepackage{cite}
\usepackage{epsfig}
\usepackage{mathrsfs}
\usepackage{eufrak}
\usepackage{epsf}
\usepackage{theorem}

\newtheorem{thm}{Theorem}

\begin{document}
\pagenumbering{arabic}
\title{\textbf{On The Capacity Of Time-Varying Channels With Periodic Feedback}}
\author{Mehdi Ansari Sadrabadi, Mohammad Ali Maddah-Ali and Amir K. Khandani \\
\small Coding \& Signal Transmission Laboratory (www.cst.uwaterloo.ca)\\
Dept. of Elec. and Comp. Eng., University of Waterloo\\ Waterloo, ON, Canada, N2L 3G1 \\
Tel: 519-884-8552, Fax: 519-888-4338\\e-mail: \{mehdi, mohammad,
khandani\}@cst.uwaterloo.ca}
\maketitle

\begin{abstract}
The capacity of time-varying channels  with
periodic feedback at the transmitter is evaluated. It is assumed that the
channel state information is perfectly known at the receiver
and is fed back to the transmitter at the regular time-intervals. The system capacity is investigated in two cases: i) finite state Markov channel, and  
ii) additive white Gaussian noise channel with time-correlated fading.
In the first case, it is shown that the capacity is achievable by multiplexing multiple codebooks across the channel.
In the second case,  the channel capacity and the optimal adaptive coding is obtained. It is shown that the optimal
adaptation can be achieved by a single Gaussian codebook, while
adaptively allocating the total power based on the side information at the
transmitter.
\end{abstract}


\section{Introduction}
Communications theory over time-varying channels has been widely
studied from different perspectives regarding the availability of
the channel state information (CSI) at the transmitter and/or
the receiver. Communication with perfect CSI at the transmitter is
studied by Shannon in \cite{shannon}, where the capacity is expressed
as that of an equivalent memoryless channel without side information
at either the transmitter or the receiver.  Communication with
perfect CSI at the receiver is investigated,  for example, in
\cite{ozarow}. With the assumption of perfect CSI  at both the
transmitter and the receiver, the capacity of finite state Markov
channels (FSMCs) and compound channels is studied in \cite{GV} and
\cite{wolf}, respectively. In practice, the assumption of perfect CSI
is not practical due to estimation inaccuracy, limited feedback
channel capacity, or feedback delay. Communication with imperfect
side information is well investigated in the
literature\cite{caire-shamai, gelfand, sirkant, klein}.  In
\cite{caire-shamai}, the capacity of FSMCs is
evaluated based on the assumed statistical relationship of the channel
state and side information at the transmitter. The channel capacity,
when feedback delay is taken into account, is studied in
\cite{lau, vishvanatan}. The optimal transmission and feedback
strategies with finite feedback alphabet cardinality is investigated in
\cite{lau2}.

In this paper, we consider a point-to-point time-varying channel
with perfectly known CSI at the receiver. It is assumed that the channel is
constant during a channel use and varies from one channel use to the
next, based on a Markov random process. The CSI is provided at the transmitter through a
noiseless feedback link at regularly-spaced time intervals. Every $T$ channel use, the CSI of the
current channel use is fed back to the transmitter.
We obtain the channel capacity of the system and show that it is
achievable by multiplexing $T$ codebooks across the channel. 
It is worth mentioning that for FSMCs, 
the results of \cite{caire-shamai} apply directly to compute the channel capacity, if the side information at
the transmitter and receiver are jointly
stationary. However, in our model, the side
information at the transmitter is not stationary.

Adaptive transmission is an efficient technique to increase the spectral efficiency of
time-varying wireless channel by adaptively modifying  the transmission rate, power, etc.,
according to the state of the channel seen by the receiver.
Adaptive transmission, which requires accurate channel
estimates  at the receiver and a reliable feedback path between the
receiver and transmitter, was first proposed in the late 1960's
\cite{Hayes}. A variable-rate and variable-power MQAM modulation
scheme for high-speed data transmission over fading channels is
studied in \cite{goldsmith-MQAM}\cite{adaptive-goldsmith}, where the
transmission rate and power are optimized to maximize the spectral
efficiency. We utilize the introduced feedback model to
obtain the capacity of additive white Gaussian noise (AWGN) channel with time-correlated fading. 
 It is shown that the capacity is achievable using a single codebook
with adaptively allocating power based on the side information at
the transmitter. Also, the optimum power allocation is derived.

The rest of the paper is organized as follows. In Section
\ref{discrete}, the system model is described and the channel
capacity is obtained. The capacity of
time-correlated  fading channel with periodic feedback is
derived in Section \ref{Ray}. The impact of channel correlation 
and feedback error on the capacity is evaluated in Section \ref{eval}. Finally, the paper is concluded in Section \ref{con}.

Throughout this paper, upper case letters represent random
variables; lower case letters denote a particular value of the
random variable; $a_m^n$ represents the sequence $(a_m, \cdots,
a_n)$ and $a^*$ is the complex conjugate of $a$. 

\section{Markov Channel With Feedback State}\label{discrete}
We consider a channel with discrete input $X_n \in \mathcal{X}$ and
discrete output $Y_n \in \mathcal{Y}$ at time instant $n$. The
channel state  is characterized as a finite-state first order Markov process:
\begin{equation}\label{1stMarkov}
\mbox{Pr}(u_n|u_1^{n-k})= \mbox{Pr}(u_n|u_{n-k})
\end{equation}
The state process, $U_n \in \mathcal{U}$, is independent of the
channel input and channel output:
\begin{equation}
\mbox{Pr}(x_1^N, y_1^N|u_1^N)= \prod_{n=1}^N\mbox{Pr}(x_n,y_n|u_n).
\end{equation}
It is  assumed that CSI is perfectly known at the receiver.
The CSI is provided at the transmitter through a
noiseless feedback link periodically at every $T$ symbols, i.e.,  $U_1, U_{T+1},
U_{2T+1}, \cdots $ are sent over the feedback link and  instantly received at the transmitter. Assume that the
codeword length, $N$, is an integer factor of $T$ and $M \triangleq
\frac{N}{T}$. Let us define $V_i \triangleq U_{T(i-1)+1}$ for $1
\leq i \leq M$ and $\tilde{n}\triangleq \lfloor \frac{n}{T}\rfloor +1$.

\subsubsection*{Encoding and Decoding}
Assume that $W\in \mathcal{W}$ is the message to be sent by the transmitter and 
$A_w=2^{NR}$ is the cardinality of $\mathcal{W}$. A codeword of length $N$ is a sequence of the encoding function
$\varphi_n$ which maps the set of messages to the channel input
alphabets. The input codeword at time $n$  depends on the message $w$
and the CSI at the transmitter up to time $n$, i.e.
$v_1^{\tilde{n}}$,
\begin{equation}
x_n=\varphi_n(w, v_1^{\tilde{n}}).
\end{equation}
The decoding function, $\phi$, maps a received sequence of $N$
channel outputs using CSI at the receiver to the message set such
that the decoded message is $\hat{w}=\phi(y_1^N, u_1^N)$.

\begin{thm}
The capacity of a finite state Markov channel with periodic feedback is given by
\begin{eqnarray}
\frac{1}{T} \sum_{t=1}^T  \sum_{v} {\mathrm  {Pr}}(v) \max_{q_t(x|v)}
\sum_{u} P_t(u|v) I(X;Y |u, v),
\end{eqnarray}
where $T$ is the feedback period, $P_t(u|v)={\mathrm {Pr}}_{u_i|u_{i-t+1}}(u|v)$ and $q_t(x|v)$ is the
random coding probability distribution function (PDF) parametrized with subscript $t$ to reflect the dependency on time.
\end{thm}

\subsection{Achievability}
We state a result on the capacity of FSMCs, which we then apply in the proof.
It is shown that the capacity of FSMCs with perfectly known CSI, $U$, at the receiver and side information $V$ at the transmitter  is
\cite{caire-shamai}
\begin{equation}\label{cap-sham}
C=\sum_{v} \mbox{Pr}(v) \max_{q(x|v)} \sum_{u} \mbox{Pr}(u|v)
I(X;Y |u, v ),
\end{equation}
where $U$ and $V$ are jointly stationary and ergodic with joint PDF $\mbox{Pr}(U,V)$ and $V$ is a deterministic function of $U$.

We consider the channel as $T$ parallel subchannels where the
$t^{\rm{th}}$ subchannel ($1 \leq t\leq T$) occurs in time instances
$(i-1)T+t, 1 \leq i \leq M$.
Noting that the channel state of the  $t^{\rm{th}}$ subchannel
$\{U_{(i-1)T+t}\}_{i=1}^M$ and the side information at the
transmitter $\{V_i\}_{i=1}^M=\{U_{(i-1)T+1}\}_{i=1}^M$ are jointly
stationary and ergodic, we define
$P_t(u|v)=\mbox{Pr}_{u_i|u_{i-t+1}}(u|v)$ for $1 \leq t\leq T$.  Using (\ref{cap-sham}), the achievable rate of the
$t^{\rm{th}}$ subchannel is
\begin{equation}
R_t=\sum_{v} \mbox{Pr}(v) \max_{q_t(x|v)} \sum_{u} P_t(u|v) I(X;Y|u, v ).
\end{equation}
$T$ codebooks are designed corresponding to $R_t$ for $1\leq t \leq
T$ and multiplexed  across the $T$ subchannels, i.e., at time instants
$(i-1)T+t$ for $1 \leq i \leq M$, the channel inputs from the  $t^{\rm{th}}$
codebook are sent over the channel. Therefore, the achievable rate is
\begin{equation}
R= \frac{1}{T} \sum_{t=1}^T  \sum_{v} \mbox{Pr}(v) \max_{q_t(x|v)}
\sum_{u} P_t(u|v) I(X;Y |u, v ).
\end{equation}

\subsection{Converse}
In this part, we prove the converse to the capacity theorem. The proof is motivated by the proof in \cite{caire-shamai}.
From the Fano's inequality\cite{gallager}, we have
\begin{equation}\label{fano1}
H(W|Y_1^N, U_1^N) \leq P_e \log A_w + h(P_e)=N \epsilon_N,
\end{equation}
where  $P_e=\mbox{Pr}(W\neq \hat{W})$ and $\epsilon_N \to 0$ as $N\to \infty$.
\begin{eqnarray}\label{fano2}
H(W|Y_1^N, U_1^N) &= & H(W|U_1^N)- I(W;Y_1^N|U_1^N) \notag \\
&=& NR - I(W;Y_1^N|U_1^N).
\end{eqnarray}
Using (\ref{fano1}) and (\ref{fano2}), we can write
\begin{eqnarray}\label{Rineq}
R\leq \frac{1}{N} I(W; Y_1^N |U_1^N)+\epsilon_N.
\end{eqnarray}
Then we have,
{\setlength\arraycolsep{1pt}\begin{eqnarray}
\lefteqn{ I(W; Y_1^N |U_1^N) {}}  \notag \\
&=& \sum_{n=1}^N I(W;Y_n|U_1^N, Y_1^{n-1}) \notag \\
&=& \sum_{n=1}^N  H(Y_n|U_1^N, Y_1^{n-1}) - H(Y_n|U_1^N, Y_1^{n-1}, W) \notag \\
&\leq& \sum_{n=1}^N  H(Y_n|U_n, V_1^{\tilde{n}}) - H(Y_n|U_1^N, Y_1^{n-1}, W ) \notag \\
\label{converse}&\stackrel{a}{\leq}& \sum_{n=1}^N  H(Y_n|U_n, V_1^{\tilde{n}}) - H(Y_n|U_n, X_n, V_1^{\tilde{n}}) \\
\label{fano3} &=& \sum_{n=1}^N I(X_n ; Y_n|U_n, V_1^{\tilde{n}}),
\end{eqnarray}}
where ($a$) follows from the fact that the channel output is independent of the message and past channel outputs
given the state of the channel and the channel input. On the other hand, for a given $n$, we have
{\setlength\arraycolsep{-2pt}\begin{eqnarray}\label{fano4}
 \lefteqn{ I(X_n ; Y_n|U_n, V_1^{\tilde{n}}) {}} \notag \\
 &=&\sum_{u_n, v_1^{\tilde{n}}} \mbox{Pr}(u_n | v_{\tilde{n}}, v_1^{\tilde{n}-1}) \mbox{Pr}(v_1^{\tilde{n}-1} | v_{\tilde{n}} ) \mbox{Pr}(v_{\tilde{n}} )  I(X_n ; Y_n|u_n, v_1^{\tilde{n}-1}, v_{\tilde{n}}) \notag \\
&\stackrel{b}{=}& \sum_{u_n,  v_{\tilde{n}}} \mbox{Pr}(u_n | v_{\tilde{n}}) \mbox{Pr}(v_{\tilde{n}}) \sum_{ v_1^{\tilde{n}-1}} \mbox{Pr}(v_1^{\tilde{n}-1} | v_{\tilde{n}} ) I(X_n ; Y_n|u_n, v_1^{\tilde{n}-1}, v_{\tilde{n}})\notag \\
&\stackrel{c}{\leq}&\sum_{u_n,  v_{\tilde{n}}}\mbox{Pr}(u_n | v_{\tilde{n}})\mbox{Pr}(v_{\tilde{n}}) \max_{q(x_n|v_{\tilde{n}})}  I(X_n ; Y_n|u_n, v_{\tilde{n}}),
\end{eqnarray}}
where ($b$) follows from the property in (\ref{1stMarkov}), and  ($c$) results from the concavity
of mutual information with respect to the input distribution, and $q(x_n|v_{\tilde{n}})\triangleq \sum_{ v_1^{\tilde{n}-1}} \mbox{Pr}(v_1^{\tilde{n}-1} | v_{\tilde{n}} )\mbox{Pr}(x_n|v_1^{\tilde{n}})$.  Replacing $n=(\tilde{n}-1)T+t$ in (\ref{fano4}) and using (\ref{fano3}), we have
{\setlength\arraycolsep{0pt}
\begin{eqnarray}
 \lefteqn{I(W; Y_1^N|U_1^N) {}} \notag \\
 &\leq&   \sum_{\tilde{n}=1}^M \sum_{t=1}^T \sum_{v_{\tilde{n}}} \sum_{u_{(\tilde{n}-1)T+t}} \mbox{Pr}(u_{(\tilde{n}-1)T+t}| v_{\tilde{n}})\mbox{Pr}(v_{\tilde{n}}) \times \notag \\ && \max_{q(x_{(\tilde{n}-1)T+t}|v_{\tilde{n}})}  I(X_{(\tilde{n}-1)T+t} ; Y_{(\tilde{n}-1)T+t}|u_{(\tilde{n}-1)T+t}, v_{\tilde{n}}) \notag \\ \label{fano33} \\
\label{fano34} &=& M \sum_{t=1}^T  \sum_{u,v} P_t(u|v) \mbox{Pr}(v)
\max_{q_t(x|v)} I(X ; Y|u, v),
\end{eqnarray}}
where (\ref{fano34}) follows from the fact that  $\{V_i\}_{i=1}^M$ and $\{U_{(i-1)T+t}\}_{i=1}^M$ are jointly stationary and ergodic and the right-hand side of (\ref{fano33}) does not depend on $\tilde{n}$. Using (\ref{Rineq}) and (\ref{fano34}), we have
\begin{eqnarray}
R\leq \frac{1}{T}  \sum_{t=1}^T \sum_{v} \mbox{Pr}(v) \max_{q_t(x|v)}\sum_u P_t(u|v) I(X ; Y|u, v) +\epsilon_N.
\end{eqnarray}
\rightline{$\blacksquare$}

\section{Gaussian Channel}\label{Ray}
In this section, we consider a point to point transmission over a time-correlated fading channel.
It is assumed that the channel gain is constant over each channel use (symbol)
and varies from symbol to symbol, following a first order Markovian random process.  The signal at the receiver is
\begin{equation}\label{received}
r_n=h_n x_n+ z_n,
\end{equation}
where $h_n \in \mathbb{C}$ is the fading gain and $z_n$ is AWGN  with zero mean and unit variance. It is assumed that the CSI is perfectly known to the receiver. Every $T$ channel use, the instantaneous fading gain is sent to the transmitter through a noiseless feedback link, i.e., $|h_1|, |h_{T+1}|, \cdots, |h_{(M-1)T+1}|$ are fed back   
and instantly  received at the transmitter. 

 Let us  define $u_n \triangleq |h_n|^2$ for $1 \leq n \leq N$,  $v_i \triangleq |h_{(i-1)T+1}|^2$ for $1 \leq i \leq M$ and $P_t(u|v) \triangleq \mbox{Pr}_{u_i|u_{i-t+1}}(u|v)$.    The average input power is subject to the constraint $\mathbb{E}[|x_n|^2]\leq \mathcal{P}$. In the following, $\mathbb{E}_t[g(U,V)]$ denotes the expectation value over $g(u,v)$ where $U$ and $V$ have joint PDF $P_t(u,v)$.
\begin{thm}
The capacity of time-correlated fading channel with periodic feedback is 
\begin{eqnarray}\label{Gauscap}
\max_{\overline{\rho}_1  \cdots \overline{\rho}_T} \frac{1}{T} \sum_{t=1}^T  \mathbb{E}_t[\log(1+U\overline{\rho}_t(V))],
\end{eqnarray}
subject to $\frac{1}{T} \sum_{t=1}^T \mathbb{E}[\overline{\rho}_t(V)]\leq \mathcal{P}$, where $T$ is the feedback period.
\end{thm}
First, we recount some results on the capacity of single user channels, which is applied in the proof.
A general formula for the capacity of single  user channels which is not necessarily information stable or stationary is obtained in \cite{Verdu}. Consider input $X$ and output $Y$ as sequences of finite-dimensional distribution, where $Y$ is induced by $X$  via  a channel which is an arbitrary sequence of finite-dimensional
conditional output distribution from input alphabets to the output alphabets. The general formula for the channel capacity is as follows:
\begin{equation}\label{inf}
C= \sup_{X} \underline{I}(X;Y),
\end{equation}
where $\underline{I}(X;Y)$ is defined as the liminf in probability
of the normalized information density \cite{Verdu}
\begin{equation}
i_N(X_1^N;Y_1^N)=\frac{1}{N} \log \frac{\mbox{Pr}(Y_1^N | X_1^N)}{\mbox{Pr}(Y_1^N)}.
\end{equation}
Assume that the channel state information, $Q$, is available at the receiver.
Considering $Q$ as an additional output,   the channel capacity is
$C=\sup_X \underline{I}(X;Y,Q)$.  If $Q$ is not available at
 the transmitter and is consequently independent of $X$, then the capacity is \cite{Narayan}
\begin{equation}\label{capcond}
C= \sup_{X} \underline{I}(X;Y|Q),
\end{equation}
 where $\underline{I}(X;Y|Q)$ is the liminf in probability of the normalized conditional information density
\begin{equation}\label{infcond}
i_N(X_1^N;Y_1^N|Q_1^N)=\frac{1}{N} \log \frac{\mbox{Pr}(Y_1^N | X_1^N, Q_1^N)}{\mbox{Pr}(Y_1^N|Q_1^N)}.
\end{equation}
Now, we are ready to prove Theorem 2, where the proof  is motivated by the proof in \cite{caire-shamai}.

\subsection{Achievability}
Noting (\ref{received}), the processed received signal at time $n$ is
\begin{equation}\label{y}
y_n=r_n \frac{h_n^*}{|h_n|} =|h_n|x_n+z^\prime_n,
\end{equation}
where $z^\prime_n=\frac{h_n^*}{|h_n|}z_n $, which has the same distribution as $z_n$.
 The transmitter sends
\begin{equation}\label{x}
x_n=\sqrt{\rho_n(v_{\tilde{n}}) }s_n,
\end{equation}
over the channel where  $s_n$ is an i.i.d. Gaussian codebook
with zero mean and unit variance, and $\rho_n:\mathbb{R}_+ \to \mathbb{R}_+$ is the power allocation function. Using (\ref{y}) and (\ref{x}), we can write
\begin{equation}\label{newch}
y_n=\sqrt{q_n} s_n+z^\prime_n,
\end{equation}
where $q_n =\rho_n (v_{\tilde{n}})|h_n|^2= \rho_n (v_{\tilde{n}}) u_n$. Noting (\ref{newch}), we have a channel with input $S$ and output $Y$ and channel state $Q$, which is known at the receiver.  Since  $Q_1^N$ is independent of  $S_1^N$, we can use (\ref{capcond}) to obtain the achievable rate.
{\setlength\arraycolsep{1pt}\begin{eqnarray}\label{infcond}
i_N(S_1^N;Y_1^N|Q_1^N)&=& \frac{1}{N} \log \frac{\mbox{Pr}(Y_1^N | S_1^N, Q_1^N)}{\mbox{Pr}(Y_1^N|Q_1^N)} \notag \\
&\stackrel{d}{=}&\frac{1}{N} \sum_{n=1}^N\log \frac{\mbox{Pr}(Y_n | S_n, Q_n)}{\mbox{Pr}(Y_n|Q_n)} \notag \\
&=& \frac{1}{N} \sum_{n=1}^N  \left( \log(1+Q_n) + \frac{|Y_n|^2}{1+Q_n}- |Z^\prime_n|^2\right),
\end{eqnarray}}
where $(d)$ results from the fact that $S_1^N$ and ${Z^\prime}_1^N$ are i.i.d. sequences and the last line follows from the fact that $Y_n$ conditioned on $Q_n$
is  Gaussian with zero mean and variance $1+Q_n$.  Note that as $N \to
\infty$, $\frac{1}{N} \sum_{n=1}^N   \frac{|Y_n|^2}{1+Q_n}=
\frac{1}{N} \sum_{n=1}^N |Z^\prime_n|^2 =1$ with probability one.
 Therefore,  with probability one, we have
{\setlength\arraycolsep{1pt}\begin{eqnarray}\label{zim}
\lefteqn{i_N(S_1^N;Y_1^N|Q_1^N) {}} \notag \\
 &=& \frac{1}{N} \sum_{n=1}^N  \log(1+Q_n) \notag \\
&=&\frac{1}{MT} \sum_{t=1}^{T} \sum_{i=1}^M \log(1+Q_{(i-1)T+t}) \notag \\
&=&  \frac{1}{T} \sum_{t=1}^{T} \frac{1}{M} \sum_{i=1}^M \log\left(1+ U_{(i-1)T+t}\rho_{(i-1)T+t} (V_i) \right).
\end{eqnarray}}
Noting that $\{U_{(i-1)T+t}\}_{i=1}^M$ and $\{V_i\}_{i=1}^M$  are
jointly stationary and ergodic for $1\leq t \leq T$, we define $P_t(u,v)$ to be their joint PDF. We set
$\rho_{(i-1)T+t}=\overline{\rho}_t$ for $1 \leq i \leq M$ and $1\leq t \leq T$.
As $M \to \infty$ in (\ref{zim}), the sample mean converges in probability to the expectation. Therefore, the achievable rate is
\begin{equation}
R=\frac{1}{T} \sum_{t=1}^{T} \mathbb{E}_t[ \log(1+ U\overline{\rho}_{t} (V) )].
\end{equation}

\subsection{Converse}
Using (\ref{converse}),  we have
{\setlength\arraycolsep{1pt}\begin{eqnarray}\label{conv1}
I(W; Y_1^N| U_1^N) &\leq& \sum_{n=1}^N  H(Y_n|U_n, V_1^{\tilde{n}}) - H(Y_n|U_n, X_n, V_1^{\tilde{n}}) \notag \\
& \leq & \sum_{n=1}^N \mathbb{E}[\log(1+ U_n \mathbb{E}[|X_n|^2|V_1^{\tilde{n}}])].
\end{eqnarray}}
The above inequality relies on the facts that
\begin{equation}
H(Y_n|U_n, X_n, V_1^{\tilde{n}})=H(Z_n)= \log 2\pi e
\end{equation}
and
\begin{equation}\label{hh}
H(Y_n|U_n, V_1^{\tilde{n}}) \leq \mathbb{E}[\log(2\pi e(1+ U_n \mathbb{E}[|X_n|^2|V_1^{\tilde{n}}]))].
\end{equation}
The upper-bound in (\ref{hh}) is achieved if $X_n$ conditioned on $V_1^{\tilde{n}}$ has a Gaussian distribution. We set $x_n=\sqrt{f_n(v_1^{\tilde{n}})} s_n$ where $f_n:\mathbb{R}_+^{\tilde{n}} \to \mathbb{R}_+$ and $S_1^N$ is an i.i.d. Gaussian sequence with zero mean and unit variance. On the other hand,
{\setlength\arraycolsep{1pt}\begin{eqnarray}\label{conv2}
\mathbb{E}[\log(1+ U_nf_n(V_1^{\tilde{n}}))] &=&\mathbb{E}[\mathbb{E}[\log(1+ U_nf_n(V_1^{\tilde{n}}))|U_n, V_{\tilde{n}} ]] \notag \\
&\stackrel{d}{\leq}& \mathbb{E}[\log(1+ \mathbb{E}[U_n f_n(V_1^{\tilde{n}})|U_n, V_{\tilde{n}}]) ]\notag \\
&=& \mathbb{E}[\log(1+ U_n \mathbb{E}[f_n(V_1^{\tilde{n}})|V_{\tilde{n}}])],
\end{eqnarray}}
where $(d)$ follows from the concavity of the logarithm.
Let us define $\rho_n(V_{\tilde{n}})\triangleq \mathbb{E}[f_n(V_1^{\tilde{n}}))|V_{\tilde{n}}]$. By using (\ref{conv1}) and (\ref{conv2}), we obtain
{\setlength\arraycolsep{1pt}\begin{eqnarray}\label{zoz}
\lefteqn{\frac{1}{N} I(W; Y_1^N| U_1^N) {}}\notag \\
 &\leq& \frac{1}{N}\sum_{n=1}^N  \mathbb{E}[\log(1+ U_n \rho_n(V_{\tilde{n}}))] \notag \\
&=& \frac{1}{T}\sum_{t=1}^T \frac{1}{M} \sum_{i=1}^M \mathbb{E}[\log(1+ U_{(i-1)T+t}\rho_{(i-1)T+t} (V_i) )] \notag \\
&\leq& \frac{1}{T}\sum_{t=1}^T  \mathbb{E}[\log(1+  \frac{1}{M}\sum_{i=1}^M U_{(i-1)T+t}\rho_{(i-1)T+t}(V_i) )]. 
\end{eqnarray}}
Using (\ref{zoz}) and noting the fact that that $\{U_{(i-1)T+t}\}_{i=1}^M$ and $\{V_i\}_{i=1}^M$
are jointly stationary and ergodic for $1\leq t \leq T$, we can write
{\setlength\arraycolsep{1pt}\begin{eqnarray}\label{zer}
\frac{1}{N} I(W; Y_1^N| U_1^N) &\leq& \frac{1}{T}\sum_{t=1}^T \mathbb{E}_t[\log(1+  U \overline{\rho}_t(V) )],
\end{eqnarray}}
where $\overline{\rho}_t(.) \triangleq \frac{1}{M}\sum_{i=1}^{M} \rho_{(i-1)T+t}(.)$. Combining (\ref{Rineq}) and (\ref{zer}), we conclude that
\begin{eqnarray}
R\leq \frac{1}{T} \sum_{t=1}^T \mathbb{E}_t[\log(1+U \overline{\rho}_t(V))],
\end{eqnarray}
subject to $\frac{1}{T} \sum_{t=1}^T \mathbb{E}[\overline{\rho}_t(V)]\leq \mathcal{P}$.\\
\rightline{$\blacksquare$}

\textit{Remark:} In Section \ref{discrete}, we prove that the capacity of Markov channels is generally achieved by using multiple code multiplexing technique.
However, for AWGN channel with time-correlated fading, the proof relies on using one Gaussian codebook, where the symbols are adaptively scaled by the appropriate power allocation function based on the side information at the transmitter.

\section{Performance Evaluation}\label{eval}
\begin{figure}[bhpt]
\centering
\includegraphics[scale=.6,clip]{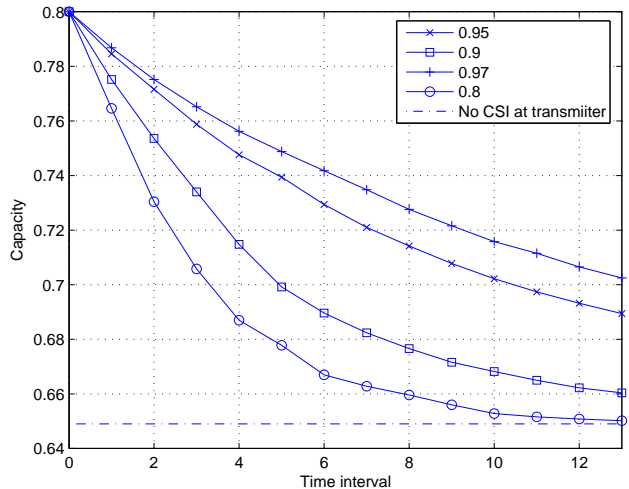}
\caption{Capacity of time-correlated Rayleigh fading channel versus  $T$
for $\mbox{SNR}=1$ and channel correlation coefficients $\alpha=0.97, 0.95, 0.9, 0.8$. The dash-dot line is the capacity with no side information at the transmitter.}
\label{fig:Tcapacity}
\end{figure}

We study the impact of the channel correlation and feedback period on
the capacity of the time-correlated Rayleigh fading channel. Let us assume that time-correlated
Rayleigh fading channel is a Markov random process with the
following PDF\cite{fading}:
\begin{eqnarray}\label{pu0}
\mbox{Pr}(u)=\left\{ \begin{array}{ll} e^{-u} & u\geq 0\\
0. & \textrm{otherwise}\end{array} \right.
\end{eqnarray}

\begin{eqnarray}
P_1(u|v)&=&\delta(v)\notag \\ 
P_t(u|v)&=&\Phi(u,v,\alpha^{t-1}),
\end{eqnarray}
where
\begin{eqnarray}\label{pdef}
\scriptstyle{ \Phi(u,v, \sigma)=\left\{ \begin{array}{ll}
\frac{1}{1-\sigma^{2}}\exp\left(-\frac{u+\sigma^{2}v}{1-\sigma^{2}}\right)\mathcal{I}_0(\frac{2\sigma
\sqrt{u v}}{1-\sigma^{2}}) & u\geq 0, \\ 0 &  \textrm{otherwise.}
\end{array} \right.}
\end{eqnarray}
In (\ref{pdef}),  $0 <\sigma <1$ describes the channel correlation
coefficient and $\mathcal{I}_0(.)$ denotes the modified
Bessel function of order zero. Noting that the capacity in (\ref{Gauscap}) is a strictly concave region of $\overline{\rho}_t, 1\leq t \leq T $, we numerically solve the 
convex optimization problem. 
In Figure \ref{fig:Tcapacity}, the capacity is depicted versus the feedback period
for various channel correlation coefficients and compared to the capacity when no CSI is available at the transmitter. 

\section{Conclusion}\label{con}
We have obtained the capacity of finite state Markov channel with
periodic feedback at the transmitter. Also, the
channel capacity and optimal adaptive coding is derived for  the time-correlated 
fading channel with periodic feedback. It is shown that the optimal adaptation can be achieved by a
single Gaussian codebook, while scaling by the appropriate power.


\begin{thebibliography}{10}

\bibitem{shannon}
{C. E. Shannon}, ``Channels with side information at the transmitter,'' {\em
  IBM J. Res. Devel.}, no.~289-293, 1958.

\bibitem{ozarow}
{L. H. Ozarow, S. Shamai, and A. D. Wyner}, ``Information theoretic
  considerations for cellular mobile radio,'' {\em IEEE Trans. Veh.
  Technology}, vol.~43, pp.~359--378, May 1994.

\bibitem{GV}
{A. Goldsmith and P. Varaiya}, ``Capacity of fading channels with channel side
  information,'' {\em IEEE Trans. Inform. Theory}, vol.~43, pp.~1986--1992,
  Nov. 1997.

\bibitem{wolf}
J.~Wolfowitz, {\em Coding {Theorems} of {Information} Theory}.
\newblock New York: Springer-Verlag, 1978.

\bibitem{caire-shamai}
{G. Caire and S. Shamai}, ``On the capacity of some channels with channel state
  information,'' {\em IEEE Trans. Inform. Theory}, vol.~45, pp.~2007 -- 2019,
  Sept. 1999.

\bibitem{gelfand}
{S. Gelfand and M. Pinsker}, ``Coding for channels with random parameters,''
  {\em Probl. Control Inform. Theory}, vol.~9, pp.~19--31, 1980.

\bibitem{sirkant}
{M. Medard and R. Srikant}, ``Capacity of nearly-decomposable markovian fading
  channels under asymmetric receiver-sender side information,'' in {\em Int.
  Symp. Inform. Theory}, p.~413, 2000.

\bibitem{klein}
{ T. E. Klein}, ``Capacity of {Gaussian} noise channels with side information
  and feedback,''
\newblock Ph.D. thesis, MIT, Feb. 2001.

\bibitem{lau}
{V. K. N. Lau}, ``Channel capacity and error exponents of variable rate
  adaptive channel coding for rayleigh fading channels,'' {\em IEEE Trans.
  Commun.}, vol.~47, pp.~1345 -- 1356, Sept. 1999.

\bibitem{vishvanatan}
{H. Viswanathan}, ``Capacity of markov channels with receiver {CSI} and delayed
  feedback,'' {\em IEEE Trans. Inform. Theory}, vol.~45, pp.~761 -- 771, March
  1999.

\bibitem{lau2}
{Vincent K. N. Lau, Youjian Liu, and Tai-Ann Chen}, ``Capacity of memoryless
  channels and block-fading channels with designable cardinality-constrained
  channel state feedback,'' {\em IEEE Trans. Inform. Theory}, vol.~50, pp.~2038
  -- 2049, Sept. 2004.

\bibitem{Hayes}
{J. F. Hayes}, ``Adaptive feedback communications,'' {\em IEEE Trans. Commun.
  Technol.}, vol.~COM-16, pp.~29--34, Feb. 1968.

\bibitem{goldsmith-MQAM}
{ Goldsmith, A.J.; Soon-Ghee Chua}, ``Variable-rate variable-power {MQAM} for
  fading channels,'' {\em IEEE Trans. Commun.}, vol.~45, pp.~1218 -- 1230, Oct.
  1997.

\bibitem{adaptive-goldsmith}
{A. Goldsmith and S. Chua}, ``Adaptive coded modulation for fading channels,''
  {\em IEEE Trans. Commun.}, vol.~46, pp.~595--602, May 1998.

\bibitem{gallager}
R.~G. Galager, {\em Information theory and Reliable communication}.
\newblock J. Wiley, New York, 1968.

\bibitem{Verdu}
{S. Verdu and T. S. Han}, ``A general formula for channel capacity,'' {\em IEEE
  Trans. Infor. Theory}, vol.~40, pp.~1147--1157, July 1994.

\bibitem{Narayan}
{A. Das and P. Narayan}, ``On the capacities of a class of finite-state
  channels with side information,'' in {\em Proc. CISS'98}, March 1998.

\bibitem{fading}
{A. N. Trofimov}, ``Convolutional codes for channels with fading,'' in {\em
  Proc. Inform Transmission}, vol.~27, pp.~155--165, Oct. 1991.

\end{thebibliography}
\end{document}